\documentclass{article}

\usepackage[dblblindworkshop, final, nonatbib]{neurips_2025}

\usepackage[utf8]{inputenc} 
\usepackage[T1]{fontenc}    
\usepackage{hyperref}       
\usepackage{url}            
\usepackage{booktabs}       
\usepackage{amsfonts}       
\usepackage{nicefrac}       
\usepackage{microtype}      
\usepackage{xcolor}         
\usepackage{subcaption}
\usepackage[tableposition=above]{caption}

\usepackage{j}  
\usepackage{wrapfig}

\title{Linear RNNs for autoregressive generation\\of long music samples}
\workshoptitle{AI for Music}

\author{%
  Konrad Szewczyk\thanks{Equal contribution.}\\
  University of Amsterdam\\
  \texttt{konrad.szewczyk@student.uva.nl}\\
  \And
  Daniel Gallo Fernández$^*$\\
  University of Amsterdam\\
  \texttt{daniel.gallo.fernandez@student.uva.nl}\\
  \AND
  James Townsend\\
  University of Amsterdam\\
  \texttt{j.h.n.townsend@uva.nl}\\
}

\begin{document}

\maketitle
\vspace{-.3cm}
\begin{abstract}
  Directly learning to generate audio waveforms in an autoregressive manner is
  a challenging task, due to the length of the raw sequences and the existence
  of important structure on many different timescales. Traditional approaches
  based on recurrent neural networks, as well as causal convolutions and
  self-attention, have only had limited success on this task. However, recent
  work has shown that deep state space models, also referred to as linear RNNs,
  can be highly efficient in this context. In this work, we push the boundaries
  of linear RNNs applied to raw audio modeling, investigating the effects of
  different architectural choices and using context-parallelism to enable
  training on sequences up to one minute (1M tokens) in length. We present a
  model, HarmonicRNN, which attains state of the art log-likelihoods and
  perceptual metrics on small-scale datasets.
\end{abstract}

\section{Introduction}
Sequence-to-sequence models have become central in artificial intelligence,
particularly following the introduction of the transformer architecture. While
initially developed for natural language processing, these models have
demonstrated utility across various domains. A notable example is computer
vision, where vision transformers are increasingly displacing traditional
convolutional neural networks. Sequence-to-sequence models require mechanisms
to exchange information along the time dimension, typically using recurrent or
self-attention layers. Recurrent layers need to compress the past into a
fixed-size state, while self-attention uses a state with size growing linearly
with the sequence length. Using \(T\) to denote sequence length, the time
complexity of recurrent layers is \(O(T)\). Self-attention, on the other hand,
is \(O(T^2)\) but is easily parallelizable at training time. However, the
quadratic complexity of self-attention becomes a serious problem at
inference (generation) time, particularly for long sequences, which commonly
occur in audio modeling.

Recent work by
\citet{guEfficientlyModelingLong2021,goelItsRawAudio2022,orvietoResurrectingRecurrentNeural2023}
has shown that recurrent neural network (RNN) layers with a \emph{linear}
recurrence can be effective for long sequence modeling, as well as enabling
training time parallelism \citep{smithSimplifiedStateSpace2022}. In this work
we propose HarmonicRNN, a sequence-to-sequence model that uses linear recurrent
layers with pooling to reduce the effective sequence length, inspired by the
SaShiMi model introduced by \citet{goelItsRawAudio2022}. We demonstrate
HarmonicRNN on autoregressive modeling of raw audio, and use multi-host context
parallelism to enable training directly on sequences up to 1M tokens (1 minute
at 16\,kHz) in length. We show that pooling is necessary in order to attain
coherent sounding samples over a long timescale, and report state of the art
log-likelihood and perceptual metrics on small audio benchmark datasets.

\section{Method}
\begin{figure*}[t!]
    \centering
    \begin{subfigure}[t]{0.29\textwidth}
        \centering
        \includegraphics[height=2.5in]{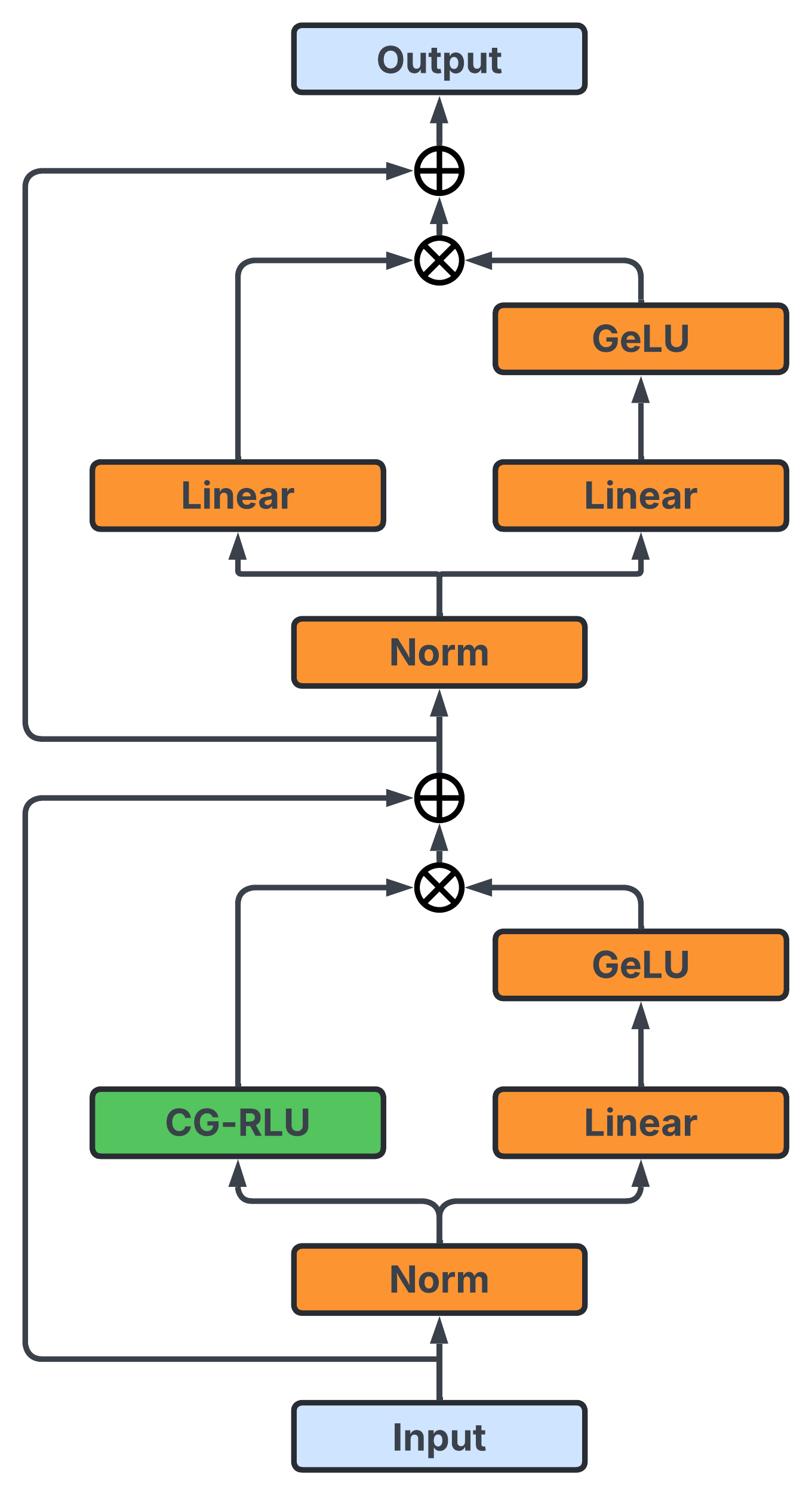}
        \caption{Temporal block.\label{fig:temp-block}}
    \end{subfigure}%
    ~
    \begin{subfigure}[t]{0.7\textwidth}
        \centering
        \includegraphics[height=2.5in]{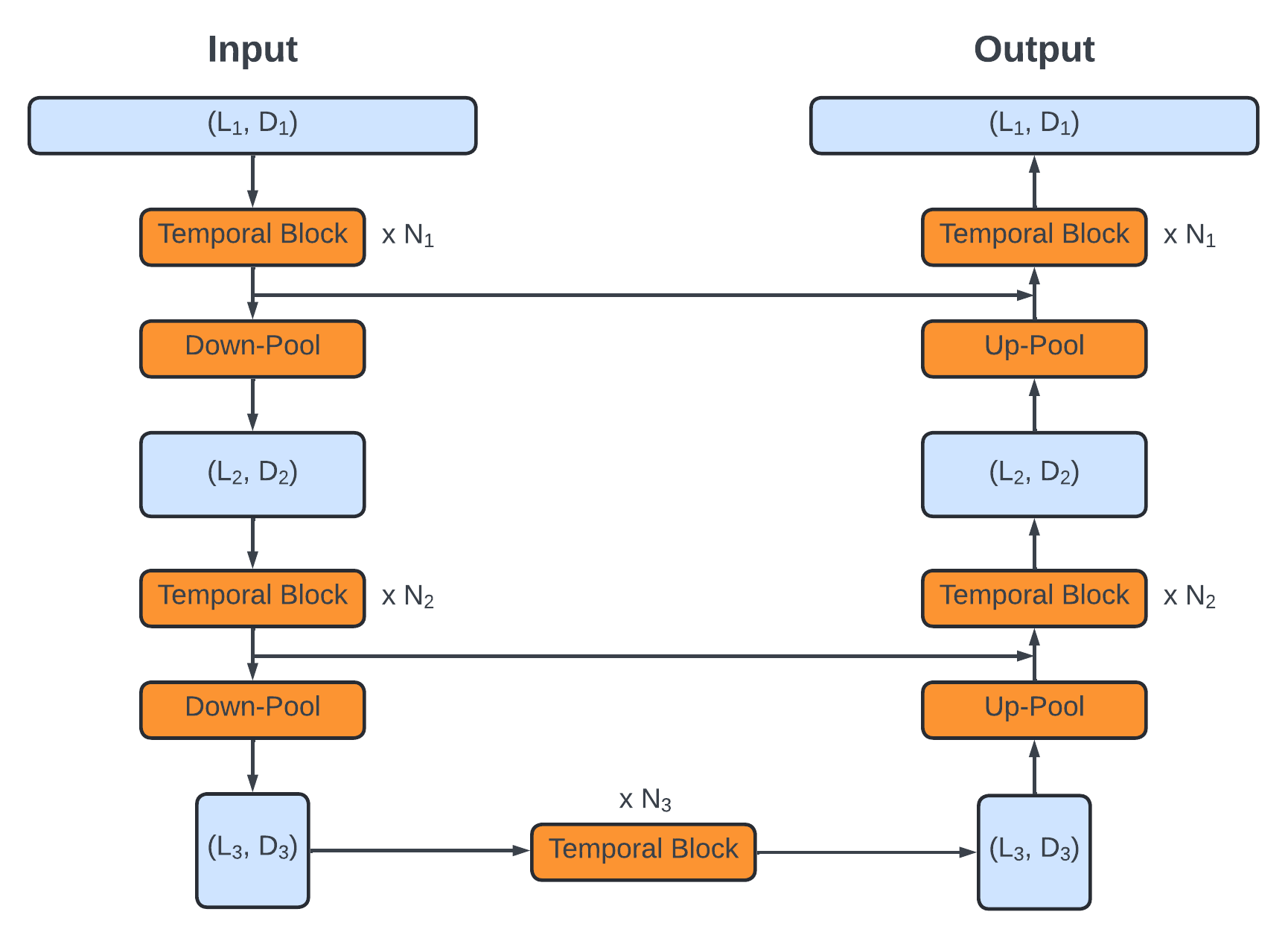}
        \caption{Overall architecture of the HarmonicRNN, here with two pooling
        layers.\label{fig:overall-arch}}
    \end{subfigure}
    \caption{Data flow graphs for the HarmonicRNN.}
\end{figure*}

We apply maximum likelihood training directly to audio data (although some
quantization preprocessing was performed, see \cref{sec:datasets}). That is,
we maximize the log-probability
\begin{equation}
    L(\theta):=\log p(x; \theta) = \sum_t \log p(x_t\given x_1,\ldots,x_{t-1}; \theta)
\end{equation}
using stochastic gradient ascent on mini-batches. The conditional probability
distributions \(p(x_t\given x_1,\ldots,x_{t-1}; \theta)\) are implemented using
a deep linear recurrent neural network (RNN), which, for each \(t\), outputs a
categorical distribution over \(x_t\) depending only on
\(x_1,\ldots,x_{t-1}\) and \(\theta\). The `linear' in linear RNN refers to the
hidden state recurrence, which in fact has the form
\begin{equation}
    h_t = a(u_t; \theta) \odot h_{t - 1} + b(u_t; \theta),
\end{equation}
where \(h_t\) and \(u_t\) are the hidden state and input, respectively, at time
\(t\); \(a\) and \(b\) are (possibly nonlinear) functions and \(\odot\) denotes
element-wise multiplication of vectors. This form allows efficient parallelized
implementation at training time using an associative scan, and has been shown
to have excellent performance
\citep{goelItsRawAudio2022,guMambaLinearTimeSequence2024,deGriffinMixingGated2024}.
We use a particular setting of \(a\) and \(b\) based on the `complex gated
linear recurrent unit' (CG-LRU), for details see
\citet{deGriffinMixingGated2024,botevRecurrentGemmaMovingTransformers2024}. For
details on the parallelized associative scan algorithm see
\citet{blellochPrefixSumsTheir1991}. The CG-LRU is wrapped in a `temporal
block', the overall structure of which is based on the transformer
architecture, with self-attention replaced by the CG-LRU, as shown in
\cref{fig:temp-block}.

On a higher level, we also enable the HarmonicRNN to model different timescales by using temporal down and up-pooling operations, with an overall architecture visualized in \cref{fig:overall-arch}. This is inspired by the SaShiMi architecture introduced in \citet{goelItsRawAudio2022}. Our model differs from SaShiMi in the following ways:

\begin{enumerate}
    \item Use of the CG-LRU as the core recurrent layer, where SaShiMi used a
      layer based on the earlier S4 architecture of
      \citet{guEfficientlyModelingLong2021}.
    \item We use strided convolutions for down-pooling and dilated (sometimes
      called transposed strided) convolutions for up-pooling. Crucially, we
      found that setting the number of feature groups greater than 1 greatly
      improved stability (see \cref{tab:groups}). This differs from SaShiMi,
      which used reshape with a dense layer, allowing dense interactions
      between input/output features.
    \item We used a non-learned embedding layer, with sinusoids of different
      periods (inspired by \cite{kingmaVariationalDiffusionModels2021}; Appendix
      C), instead of a learned embedding layer. We found that this led to
      faster training (see \cref{fig:embeddings}).
\end{enumerate}

\section{Related work}
\paragraph{State space models and linear RNNs}This work builds on recent
progress using state space models (from classical control theory) as layers in
a deep neural network. This line of work was initiated by
\citet{guEfficientlyModelingLong2021}, and further developed by
\citet{smithSimplifiedStateSpace2022, guMambaLinearTimeSequence2024,
orvietoResurrectingRecurrentNeural2023,deGriffinMixingGated2024},
among others.

\paragraph{Autoregressive generation of raw audio}
The major models proposed for this task are SampleRNN
\citep{mehriSampleRNNUnconditionalEndtoEnd2017}; WaveNet and its derivatives
\citep{vandenoordWaveNetGenerativeModel2016, oordParallelWaveNetFast2018,
kalchbrennerEfficientNeuralAudio2018};
and recently SaShiMi \citep{goelItsRawAudio2022,guMambaLinearTimeSequence2024}.
SampleRNN and WaveNet use (nonlinear) RNNs and causal convolutions,
respectively. SaShiMi was the first model where state space
models---effectively equivalent to linear RNNs
\citep{orvietoResurrectingRecurrentNeural2023}---were applied to audio
modeling, with state of the art results.

\section{Experiments}\label{sec:experiments}
We performed ablation experiments to evaluate various aspects of our model in terms of log-likelihood, perceptual metrics and inference speed.
\subsection{Datasets}\label{sec:datasets}
We use the same three datasets and train / test splits from
\citet{goelItsRawAudio2022}. All of them are audio recordings sampled at
16\,kHz with 16-bit linear PCM encoding.
\begin{enumerate}
    \item \textbf{SC09}
      \citep{donahueAdversarialAudioSynthesis2018,wardenSpeechCommandsDataset2018},
      which consists of 1-second (16k token) recordings of the spoken digits
      zero to nine (similar to MNIST, but with audio instead of images). For
      this dataset, we also compute the FID and IS scores using the SaShiMi
      repo \citep{goelItsRawAudio2022}.
    \item \textbf{Beethoven} \citep{mehriSampleRNNUnconditionalEndtoEnd2017},
      which contains 8-second (128k token) recordings of Beethoven’s piano
      sonatas.
    \item \textbf{YouTubeMix} \citep{DeepsoundprojectSamplernnpytorch2025},
      which contains 1-minute (960,512 token) recordings of piano playing.
\end{enumerate}
Following \citet{goelItsRawAudio2022}, we apply µ-law encoding to SC09 and
YouTubeMix, and linear encoding to Beethoven, and quantize to an unsigned 8-bit
representation. All of the datasets are available on HuggingFace datasets:
\href{https://huggingface.co/datasets/krandiash/beethoven}{Beethoven},
\href{https://huggingface.co/datasets/krandiash/youtubemix}{YouTubeMix}, and
\href{https://huggingface.co/datasets/krandiash/sc09}{SC09}.

\subsection{Baseline model and training setup}
We used the JAX, Flax and Optax libraries to implement our experiments and ran
them on TPU v4-8 and v3-128 devices. For training on the minute-long samples in
YouTubeMix, we needed to use multi-host training, because of the need for more
TPU device memory. We used the custom CG-LRU TPU kernels from the
\href{https://github.com/google-deepmind/recurrentgemma}{RecurrentGemma}
repository, which support context parallelism (parallelism over the sequence
axis) to split the model across devices.

\paragraph{Model} We use the model structure visualized in
\cref{fig:overall-arch}, with four levels of down-pooling. The down-pooling
factors (from outer to inner) are [2, 4, 4, 5], meaning that at the innermost
layer, the sequence length is \(2\times 4\times 4\times 5 = 160\) times smaller
than the data sequence. The number of temporal mixing blocks between the
pooling layers is 4 at every level, in both the down-pooling and up-pooling
parts of the network, leading to a total of 36 temporal blocks. We use a
feature width of 128, and an RNN hidden dimension of 256. This setup gives a
model with 7.3M parameters.

\paragraph{Training}
We use the AdamW optimizer \citep{loshchilovDecoupledWeightDecay2018} with a
learning rate of 0.002 and a weight decay of 0.0001. All other hyperparameters
are set to the default values in Optax: \(\beta_1 = 0.9\), \(\beta_2 = 0.999\).
We used a constant learning rate schedule after 1,000 steps warm-up, batch size
32 and 500 training epochs. We evaluated on the test sets. We used exponential
moving average (EMA) for the model weights, with rate equal to 0.999.

\begin{wrapfigure}{R}{0.40\textwidth}
  \vspace{-.5cm}
  \captionsetup{font=footnotesize}
  \centering
    \includegraphics[width=0.38\textwidth]{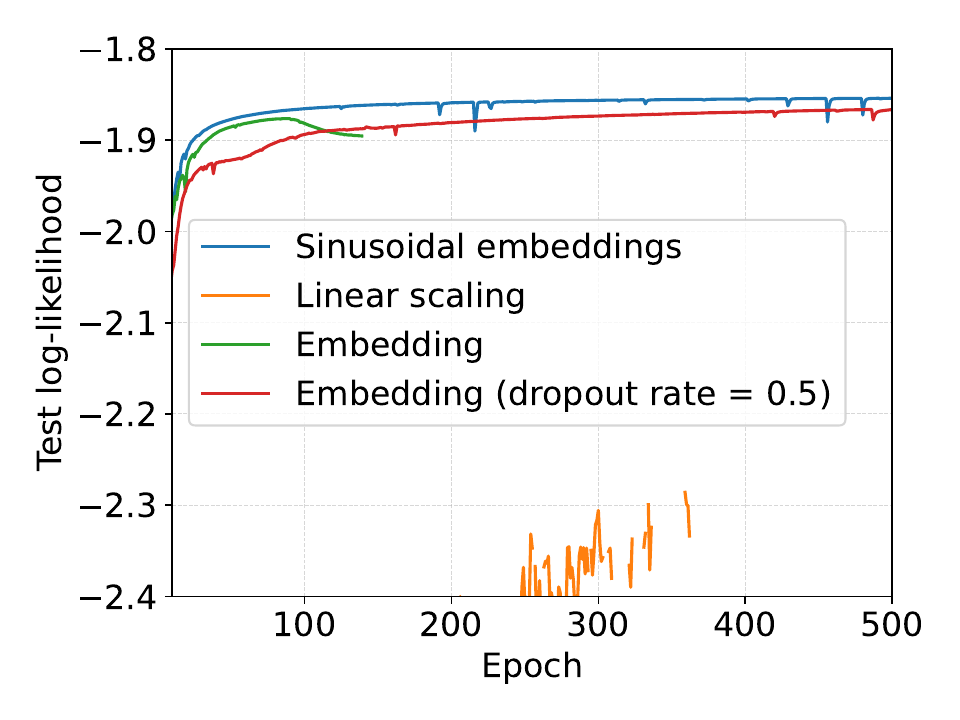}
  \caption{Training curves with different embedding
  methods (SC09 dataset).}\label{fig:embeddings}
\end{wrapfigure}

\subsection{Effect of input embedding}
As described in \cref{sec:datasets}, the model's input is an array of 8-bit
unsigned integers. We tried four approaches to embed data before feeding it to
the neural network; \cref{fig:embeddings} shows a comparison of training
curves. `Linear scaling' was simply scaling the data into the real interval
[-1, 1] (this technique is used by e.g.\ WaveNet). The sinusoidal embeddings
were sinusoids of different frequencies applied to the input integers, based on
\citet[][Appendix C]{kingmaVariationalDiffusionModels2021}. Finally, we tried a
standard embedding layer, as is commonly used in transformer architectures,
with and without dropout. We found that the sinusoidal embeddings consistently
performed better, with significantly faster convergence and better final loss
value.

\subsection{Effect of pooling}
\begin{table}[h]
  \captionsetup{font=scriptsize}
  \scriptsize
\parbox{.34\linewidth}{
  \begin{tabular}[t]{lcc}
\toprule
\# Groups&\# Params&Test NLL \(\downarrow\)\\
\midrule
1 (dense)     &7.8M&1.956\\
4             &7.4M&\textbf{1.848}\\
128 (diagonal)&7.3M&1.854\\
\bottomrule
\end{tabular}
\caption{Effect of number of feature groups in pooling convolutions on total
parameter count and test NLL (SC09 dataset).}
\label{tab:groups}}
\hfill
\parbox{.60\linewidth}{
\centering
\begin{tabular}[t]{cccccc}
\toprule
Pooling&Train speed&Infer throughput&NLL \(\downarrow\)&FID \(\downarrow\)&IS \(\uparrow\)\\
\midrule
Baseline   &\textbf{11.5 epoch/h}& \textbf{103\,ktok/s}&1.854&\textbf{0.46}&\textbf{6.46}\\
1 pool lyr.&6.7 epoch/h          & 55\,ktok/s&1.853         &6.85&2.03         \\
No pooling &6.1 epoch/h          & 46\,ktok/s&\textbf{1.852}&2.95&3.33         \\
\bottomrule
\end{tabular}
\caption{Effect of pooling configuration on training and inference speed;
  NLL, FID and IS performance metrics. Total layer count is held constant. The
  model with 1 pooling layer has pooling factor 2 with 24 temporal blocks
  outside the pooling and 12 within.}\label{tab:pooling} }
\end{table}
We measured the effect on performance of the group count in the pooling
convolutions (\Cref{tab:groups}), and the effect on training and inference
time, as well as other metrics, of removing pooling altogether
(\Cref{tab:pooling}). For the group count, we found that the best performing in
negative log-likelihood (NLL) was 4 groups, though this had slightly more
parameters and had a slower training runtime than the fully diagonal
convolution. The effect of reducing or removing pooling is to significantly
reduce training and inference speed, very slightly decrease NLL, and to worsen
significantly the perceptual metrics Frechét Inception Distance (FID) and
Inception Score (IS). We used the recently developed Scanagram library to
implement efficient inference \citep{Scanagram2025}. Inference batch size and
sequence length were the same as during training.

\subsection{Comparison to prior state of the art}

\begin{wraptable}[9]{R}{0.58\textwidth}
  \vspace{-.3cm}
  \captionsetup{font=scriptsize}
  \scriptsize
  \centering
  \begin{tabular}{lcccc}
    \toprule
    Model                      &\# Params&Test NLL \(\downarrow\)&FID \(\downarrow\)&IS \(\uparrow\)\\
    \midrule
    HarmonicRNN baseline       &7.3M     &1.854                  &\textbf{0.46}     &\textbf{6.46}\\
    HarmonicRNN 4 conv grps.   &7.4M     &\textbf{1.848}         &--                &--           \\
    SaShiMi                    &5.8M     &1.873                  &1.99              &5.13         \\
    Mamba                      &6.1M     &1.852                  &0.94              &6.26         \\
    \bottomrule\\
  \end{tabular}

  \caption{Comparison of our models with the prior state-of-the-art (SC09
    dataset). Note that
  the results for SaShiMi are those reported in
\citet{guMambaLinearTimeSequence2024}, which differ from those in the original
SaShiMi paper \citep{goelItsRawAudio2022}.}\label{tab:sota}
\end{wraptable}
In \cref{tab:sota}, we compare our best configurations with SaShiMi and Mamba
(the prior state of the art) on the SC09 dataset.  Versions of HarmonicRNN
achieved state of the art results on all metrics.  \Cref{tab:other-datasets}
compares with SaShiMi on the other two datasets.  Unlike SaShiMi, we trained on
full minute-long YouTubeMix sequences.  In the supplementary material we
provide eight minute-long, non-cherry-picked samples, which compare favourably
with the SaShiMi samples at
\url{https://hazyresearch.stanford.edu/sashimi-examples/#music}.

\section{Conclusion}
\begin{wraptable}[5]{r}{0.4\textwidth}
  \vspace{-.4cm}
  \captionsetup{font=scriptsize}
  \scriptsize
  \centering
  \begin{tabular}{lcc}
    \toprule
    Model      &Beethoven \(\downarrow\)&YouTubeMix \(\downarrow\)\\
    \midrule
    HarmonicRNN&\textbf{0.915}          &1.324                        \\
    SaShiMi    &0.946                   &\textbf{1.294}               \\
    \bottomrule\\
  \end{tabular}

  \caption{Comparison of our HarmonicRNN test NLL with the prior
    state of the art on Beethoven and YouTubeMix.\label{tab:other-datasets}}
\end{wraptable}
We have presented initial results for HarmonicRNN, a deep linear RNN designed
to generate music in an autoregressive manner. The model shows promising
performance and we look forward to future work scaling the approach up to more
challenging audio generation tasks.

\section{Acknowledgments}
James Townsend acknowledges funding from the Dutch Research Council (NWO) under
Veni project VI.Veni.212.106, and the European Commission under MSCA project
NNESCI. This work was done as part of an MSc thesis project by the two first
authors. We would like to thank Jan-Willem van de Meent for acting as examiner
for the project and for helpful discussions.

\printbibliography

\end{document}